\documentclass[amsmath,amssymb,aps,floats,amsfonts,notitlepage,superscriptaddress,nofootinbib,prl,a4paper,twocolumn,longbibliography]{revtex4-1}

\usepackage[utf8]{inputenc}
\usepackage[]{graphicx}
\usepackage{hyperref}
\usepackage{url}
\usepackage{color}
\usepackage[usenames,dvipsnames,svgnames,table]{xcolor}
\usepackage{multirow}
\usepackage{ulem}
\usepackage[scr=boondoxo,scrscaled=1.05]{mathalfa}
\usepackage{tensor}
\usepackage{booktabs}

\font\ec=ecrm0800 at 11pt
\def\th{\hbox{\ec\char'336}}
\def\edth{\hbox{\ec\char'360}}

\newcommand{\mb}{{\bar{m}}}

\newcommand{\beq}{\begin{equation}}
\newcommand{\eeq}{\end{equation}}

\newcommand{\zetac}{\bar{\zeta}}
\newcommand{\cH}{\mathcal{H}}

\newcommand{\cD}{\mathcal{D}}
\newcommand{\cL}{\mathcal{L}}
\newcommand{\LieT}{\pounds_{T}}

\begin{document}

\title{Gravitational perturbations of rotating black holes in Lorenz gauge}

\author{Sam R.~Dolan}
\affiliation{
Consortium for Fundamental Physics,
School of Mathematics and Statistics,
University of Sheffield,
Hicks Building,
Hounsfield Road,
Sheffield S3 7RH, 
United Kingdom}

\author{Chris Kavanagh}
\affiliation{
Max Planck Institute for Gravitational Physics (Albert Einstein Institute),
Am M\"uhlenberg 1,
Potsdam 14476,
Germany}

\author{Barry Wardell}
\affiliation{
School of Mathematics and Statistics,
University College Dublin,
Belfield,
Dublin 4,
Ireland
}

\begin{abstract}
Perturbations of Kerr spacetime are typically studied with the Teukolsky formalism, in which a pair of gauge invariant components of the perturbed Weyl tensor are expressed in terms of separable modes that satisfy ordinary differential equations. However, for certain applications it is desirable to construct the full metric perturbation in the Lorenz gauge, in which the linearized Einstein field equations take a manifestly hyperbolic form. Here we obtain a set of Lorenz-gauge solutions to the linearised vacuum field equations on Kerr-NUT spacetimes in terms of homogeneous solutions to the spin-$2$, spin-$1$ and spin-$0$ Teukolsky equations. We also derive Lorenz-gauge completion pieces representing mass and angular momentum perturbations of Kerr spacetime. 
\end{abstract}

\maketitle

The Kerr spacetime \cite{Kerr:1963ud} is a fundamental vacuum solution of Einstein's field equations which provides a mathematical description of the vast number of rotating black holes in our universe. Key questions on black hole stability, cosmic censorship, and gravitational-wave generation are addressed via {\it black hole perturbation theory} \cite{Pound:2021qin}, in which Kerr's solution sets the stage for the dynamics of scalar, spinor, electromagnetic and gravitational field perturbations playing out on a curved background.

The spacetime possesses obvious time-translation and axial symmetries, but also a `hidden' symmetry encoded in a conformal Killing-Yano tensor \cite{Frolov:2017kze}. This symmetry, which is closely related to the existence of a doubled pair of principal null directions (i.e. Petrov type D), underpins some remarkable results including (i) Liouville-integrability for the geodesic equations \cite{Carter:1968rr}; (ii) decoupling and separability of certain Bianchi identities, allowing the perturbed Weyl scalars $\Psi_0$ and $\Psi_4$ to be expressed as a sum of modes governed by second-order ordinary differential equations \cite{Teukolsky:1972my,Teukolsky:1973ha}; and (iii) a complete separation of variables for massive scalar \cite{Brill:1972xj}, spinor \cite{Chandrasekhar:1976ap} and vector fields \cite{Frolov:2018ezx}. Exploitation of the hidden symmetry in $(n+1)$-dimensional Kerr-NUT-(A)dS contexts is ongoing \cite{Krtous:2018bvk,Lunin:2019pwz,Houri:2019dsd,Houri:2019lnu}.

A key result from 1975 is that a metric perturbation $h_{\mu \nu}$ can be constructed from a spin-$2$ scalar Hertz potential $\psi$ in such a way as to satisfy the linearized Einstein equations on the Kerr spacetime \cite{Chrzanowski:1975wv, Wald:1978vm, Stewart:1978tm, Kegeles:1979an}. The metric perturbation so obtained is in a radiation gauge (or light-cone gauge \cite{Jackson:2001ia}), such that $h_{\mu \nu} \ell^\nu = 0$, where $\ell^\nu$ is a principal null direction.
In the presence of sources, the construction generically leads to non-isotropic particle singularities and extended gauge discontinuities in the metric perturbation \cite{Barack:2001ph,Ori:2002uv,Keidl:2010pm,Pound:2013faa}. This is an impediment to extending perturbation theory to second order, because the source terms at second order are derived from the metric perturbation at first order \cite{Pound:2017psq}. By contrast, in the Lorenz gauge
$h_{\mu \nu}$ is expected to be free from extended gauge discontinuities.

A metric perturbation $h_{\mu \nu}$ satisfying 
\beq
\nabla_\nu \widehat{h}^{\mu \nu} = 0, \label{eq:Lorenz-gauge}
\eeq
is said to be in Lorenz gauge, also known as {\it harmonic} or {\it de Donder} gauge. 
Here $\widehat{h}_{\mu \nu} = h_{\mu \nu} - \frac12 g_{\mu \nu} h$ is the trace-reversed metric perturbation, $h = \tensor{h}{^\mu _\mu}$ is its trace, and $\nabla_\mu$ denotes the covariant derivative on the background metric $g_{\mu\nu}$. 
Imposing the Lorenz-gauge condition on the linearized Einstein equations leads to the (manifestly hyperbolic) Lichnerowicz tensor wave equation,
\begin{align}
\Box \widehat{h}_{\mu \nu} + 2 \tensor{R}{_\mu ^\sigma _\nu ^\lambda} \widehat{h}_{\sigma \lambda}  = -16 \pi T_{\mu \nu} ,
\label{Lorenz-field-equation}
\end{align}
where $T_{\mu \nu}$ is the stress-energy tensor of matter sources, and $\tensor{R}{_{\mu \sigma \nu \lambda}}$ is the Riemann tensor of the background spacetime which we take to be Ricci-flat ($R_{\mu \nu} = 0$).

The gravitational self-force (GSF) programme addresses the challenge of modelling Extreme Mass-Ratio Inspirals for gravitational wave detectors. GSF calculations are naturally formulated and conducted in Lorenz gauge \cite{Mino:1996nk, Barack:2005nr, Barack:2007tm, Berndtson:2007gsc, Poisson:2011nh, Dolan:2012jg, Akcay:2013wfa, Miller:2020bft}. On Schwarzschild spacetime, a Lorenz-gauge formulation at first order \cite{Miller:2020bft} is an essential ingredient in the recent calculation of the gravitational-wave flux at second order in the mass ratio \cite{Warburton:2021kwk}. Lacking a separable solution of the Lorenz-gauge equations on Kerr spacetime in the literature (see Ref.~\cite{Whiting:2005hr} for discussion), recent focus has shifted to constructing second-order perturbations in sufficiently-regular gauges \cite{Campanelli:1998jv, Lousto:2002em, Pound:2017psq, Green:2019nam, Loutrel:2020wbw, Ripley:2020xby, Toomani:2021jlo}.

In the context of electromagnetism, a vector potential $A^\mu$ is said to be in {\it Lorenz gauge} if it satisfies $\nabla_\mu A^\mu = 0$. Imposing the Lorenz gauge condition renders the Maxwell field equation into a wave equation, $\Box A^\mu = j^\mu$.
Recent work \cite{Lunin:2017drx,Krtous:2018bvk,Dolan:2019hcw,Houri:2019dsd,Houri:2019lnu,Lunin:2019pwz,Wardell:2020naz} has identified a separable method for obtaining solutions to Maxwell's equations in Lorenz gauge on spacetimes that include Kerr. In this work, we show that a similar approach may also be applied in the context of Lorenz-gauge gravitational perturbations, by obtaining a set of solutions for the Lorenz-gauge equations \eqref{Lorenz-field-equation} on Kerr spacetime in the absence of sources ($T_{\mu \nu} = 0$) for the first time.

\textit{Preliminaries.---}
The Kerr metric can be written in terms of a null tetrad,
\beq
g^{\mu \nu} = - 2 \, l^{(\mu} n^{\nu)} + 2 \, m^{(\mu} \bar{m}^{\nu)},
\eeq
where $l^{\mu}$ and $n^{\nu}$ are aligned with the principal null directions, $m^{\mu}$ is a complex null vector and $\bar{m}^{\nu}$ is its complex conjugate.
In Boyer-Lindquist coordinates $\{ t, r, \theta, \phi \}$, the Kinnersley null tetrad is $l^\mu = l_+^\mu$, $n^\mu = - \frac{\Delta}{2\Sigma} l_-^\mu$, $m^\mu = \frac{1}{\sqrt{2} \zetac} m_+^\mu$ and $\bar{m}^\mu = (m^\mu)^\ast = \frac{1}{\sqrt{2} \zeta} m_-^\mu$, with
\begin{subequations}
\begin{align}
l^\mu_\pm &= \left[ \pm (r^2+a^2) / \Delta, 1, 0, \pm a / \Delta \right] , \\ 
m^\mu_{\pm} &= \left[ \pm i a \sin \theta, 0, 1, \pm \, i \, / \sin \theta \right],
\label{eq:tetrad2}
\end{align}
\end{subequations}
where $\Delta = r^2 - 2Mr + a^2$, $\Sigma = \zeta \zetac = r^2 + a^2 \cos^2 \theta$ and
\beq
\zeta = r - i a \cos \theta .   \label{eq:zeta-def}
\eeq
The parameters $M$ and $a$ represent the mass and specific angular momentum of the black hole.

In the absence of sources, the spin-$2$ perturbed Weyl scalars satisfy the homogeneous Teukolsky equations  \cite{Teukolsky:1972my,Teukolsky:1973ha,Chandrasekhar:1985kt} (see \cite{Pound:2021qin} for a review with conventions consistent with those used here), $\mathcal{O}\Psi_0 = 0 = \mathcal{O}' \Psi_4 \equiv \zeta^{-4} \mathcal{O} \zeta^{4} \Psi_4$. The Teukolsky equations admit a separation of variables: working with the Kinnersley tetrad and inserting the ansatz $\zeta^4 \Psi_4 = R_{-2}(r) S_{-2}(\theta) e^{-i \omega t + i m \phi}$ yields
\beq
\mathcal{O}' \Psi_4 = \zeta^{-4} \left[\Delta \cD^\dagger_{-1} \cD + \cL_{-1} \cL_2^\dagger - 6 i \omega \zetac \right] (\zeta^4 \Psi_4)  = 0 , \label{eq:Teuk}
\eeq
where the directional derivatives are $\cD \equiv l_+^\mu \partial_\mu$, $\cD^\dagger \equiv l_-^\mu \partial_\mu$, $\cL^\dagger = m_+^\mu \partial_\mu$, $\cL = m_-^\mu \partial_\mu$ with $\cD_n = \cD + n (\partial_r \Delta) / \Delta$ and $\cL_n = \cL + n \cot \theta$. The functions $R_{-2}(r)$ and $S_{-2}(\theta)$ therefore satisfy a set of decoupled ordinary differential equations. A similar result also holds for $\Psi_0$.

There is substantial {\it gauge freedom} in perturbation theory, linked to the freedom 
to make an infinitesimal coordinate transformation $x^\mu \rightarrow x^\mu + \epsilon \, \xi^\mu$, where $\epsilon=1$ is an order-counting parameter. Under such a transformation, a tensor field $\mathrm{T} = T + \epsilon \, \delta T$ changes at perturbative order as
$
 \mathrm{T}  \rightarrow  T + \epsilon \left( \delta T - \pounds_{\xi} T \right) + O(\epsilon^2)
$, 
where $\pounds_{\xi}$ denotes the Lie derivative along the gauge vector $\xi^{\mu}$. Applying this rule to the perturbed metric $\mathrm{g}_{\mu \nu} = g_{\mu \nu} + \epsilon \, h_{\mu \nu}$ yields a transformation law for the metric perturbation $h_{\mu \nu}$ under a change of gauge, namely, $h_{\mu \nu} \rightarrow h_{\mu \nu} - 2 \xi_{(\mu ; \nu)}$, where a semi-colon denotes the covariant derivative and parentheses indicate symmetrization over the indices.

On a vacuum black hole background ($R_{\mu \nu} = 0$), the perturbed Ricci tensor $\delta R_{\mu \nu}$ is gauge-invariant at linear order (as $\pounds_{\xi} R_{\mu \nu} = 0$). Consequently, any pure-gauge metric perturbation $h_{\mu \nu} = - 2 \xi_{(\mu ; \nu)}$ satisfies the vacuum field equations; furthermore if the vector satisfies $\Box \xi^\mu = 0$ then $h_{\mu \nu}$ is in Lorenz gauge and the metric perturbation satisfies Eq.~\eqref{Lorenz-field-equation} with $T_{\mu \nu} = 0$.

In principle, given a vacuum metric perturbation $h_{\mu \nu}$, one may
apply a gauge transformation to transform it to Lorenz gauge, such that
\beq
h^{L}_{\mu \nu} \equiv h_{\mu \nu} - 2 \xi_{(\mu ; \nu)} \label{eq:hLorenz}
\eeq
satisfies Eq.~(\ref{eq:Lorenz-gauge}). It follows that the gauge vector $\xi^\mu$ must satisfy a sourced wave equation,
\beq
\Box \xi^\mu = \nabla_\nu \widehat{h}^{\mu \nu}. \label{eq:gauge-source}
\eeq

\textit{Reconstruction of Lorenz gauge solutions from scalar potentials.---}
Our main result is that one can construct solutions to the Lorenz gauge equations from separable solutions of the Teukolsky equation. These solutions are divided into scalar (spin-$0$), vector (spin-$1$), and tensor (spin-$2$) type, alongside ``completion'' pieces \cite{Merlin:2016boc,vanDeMeent:2017oet} associated in the Kerr case with infinitesimal changes in the mass and angular momentum of the black hole. 
In the absence of sources, the spin-$0$ and spin-$1$ perturbations are pure-gauge modes. In the presence of sources, we anticipate that solutions of all types ($s=0$, $1$, $2$) will be required to construct a physical solution that is free from gauge discontinuities, as is found to be the case on Schwarzschild spacetime \cite{Berndtson:2007gsc}.

\textit{Spin-2 solutions.---}
To obtain Lorenz gauge solutions derived from spin-$2$ scalars, we start with the ingoing radiation-gauge solution of Chrzanowski (Ref.~\cite{Chrzanowski:1975wv}, Table I) and seek a transformation to Lorenz gauge. Chrzanowski's solution can be expressed in covariant form as \cite{Aksteiner:2016pjt}
\begin{equation}
h_{\mu \nu} = -\frac12 \nabla_\beta \left[ \zeta^{4} \nabla_\alpha \left( \zeta^{-4} \tensor{\cH}{_{(\mu} ^\alpha _{\nu)} ^\beta} \right) \right] \label{eq:hrad}
\end{equation}
where
\begin{equation}
\cH^{\mu \alpha \nu \beta} =  4 \psi \,  l^{[\mu} m^{\alpha]} l^{[\nu} m^{\beta]},
\end{equation}
and where $\psi$ is a spin-weight $-2$ potential. In the absence of sources it satisfies a homogeneous $s=-2$ Teukolsky equation, $\mathcal{O} \psi = 0$.

The metric perturbation in Eq.~\eqref{eq:hrad} is manifestly trace-free ($h = 0$). The inclusion of $\zeta^4$ is required in order to satisfy the linearised Einstein equation but violates the Lorenz gauge condition; without it the metric perturbation would automatically satisfy the Lorenz gauge condition but not the linearised Einstein equation \cite{Stewart:1978tm}. Finally, in order to obtain a real metric perturbation that generates a physical Weyl tensor one typically adds the complex conjugate of this metric perturbation; for now we omit the complex conjugate and will return to it later.

We now seek to transform $h_{\mu \nu}$ to Lorenz gauge by solving Eq.~\eqref{eq:gauge-source}, while preserving the trace-free condition. That is, we seek a gauge vector $\xi^\mu$ satisfying
\beq
\Box \xi^\mu = -j^\mu \equiv \nabla_{\nu} h^{\mu \nu} , \quad \quad \nabla_\mu \xi^\mu = 0 .
\eeq
This we recognise as a well-formed electromagnetic field equation in (vector) Lorenz gauge. The effective four-current $j^\mu$ is divergence-free ($\nabla_\mu j^\mu = 0$) by virtue of the fact that $h^{\mu \nu}$ in Eq.~\eqref{eq:hrad} satisfies $\nabla_{\mu} \nabla_{\nu} h^{\mu \nu} = 0$. 
The above becomes clearer when written in terms of forms: 
\beq
\delta \mathrm{d} \xi = j , \quad \quad \delta \xi = 0, \quad \quad \delta j = 0 . \label{eq:forms1}
\eeq
Here $\mathrm{d}$ is the exterior derivative, $\delta = {}^\star \mathrm{d} {}^\star$ is the coderivative, ${}^\star$ is the Hodge dual operation, $\Box \xi = \mathrm{d} \delta \xi - \delta \mathrm{d} \xi$ on a Ricci-flat spacetime, and a key identity is $\mathrm{d} \mathrm{d} = 0 = \delta \delta$.

By Poincar\'e's lemma, a divergence-free vector is locally the coderivative of a (non-unique) two-form. A short calculation establishes that $j = \delta J$, that is, $j^\mu = \nabla_\nu J^{\mu \nu}$ with the two-form  
\beq
J^{\mu \nu} = \nabla_\beta\left[ U_\alpha \cH^{\beta\alpha\mu\nu}\right] = \frac{ \sqrt{2}}{\Sigma} l^{[\mu} m^{\nu]} \left[\cL_2^\dagger - i a \sin \theta \cD \right] \psi ,
\eeq
where $U_a = - \nabla_a \ln \zeta$. 
Equation \eqref{eq:forms1} can be written as $\delta (\mathrm{d} \xi - J + {}^\star \mathrm{d} \varsigma ) = 0$, where $\varsigma$ is an arbitrary vector field (i.e.~a gauge vector of the third kind \cite{Cohen:1974cm}). The recent work of Green {\it et al.}~\cite{Green-talk-2021,Green-paper} suggests the ansatz 
\beq
\xi = \zeta^2 \delta H - \mathrm{d} \chi,  \label{eq:xi-ansatz}
\eeq 
where $H$ is a two-form and $\chi$ is a scalar; and we choose the gauge vector of the third kind to be $\varsigma = - i \zeta^2 \delta H$ so that the field equation becomes \cite{Green-Toomani} 
\beq
\delta \left( (1 - i {}^\star) \mathrm{d} \zeta^2 \delta H - J \right) = 0 .
\eeq
The operator $\mathrm{d} \zeta^2 \delta$ generates decoupled equations for the three anti-self-dual degrees of freedom in the two-form $H$ \cite{Green-talk-2021}; and the operator $(1 -  i {}^\star)$ annihilates the self-dual components of the equation \cite{Mustafa:1987hertz,Green-talk-2021}. The ansatz 
\beq
H^{\mu \nu} = \frac{\sqrt{2}}{\zeta} l^{[\mu} m^{\nu]} \alpha \label{eq:Hansatz}
\eeq
then leads to a single decoupled second-order equation,
\beq
\left( \Delta \cD^\dagger \zeta^2 \cD + \cL \zeta^2 \cL_1^\dagger \right) \alpha = - \zeta \left(\cL_2^\dagger - i a \sin \theta \cD \right) \psi . \label{eq:Heqn}
\eeq
Assuming harmonic time dependence $e^{- i \omega t}$ for $\psi$, and by application of the vacuum Teukolsky equation \eqref{eq:Teuk}, we find that Eq.~\eqref{eq:Heqn} has an elementary solution,
\beq
\alpha = - \frac{1}{6 i \omega \zeta} \cD \cL_2^\dagger \psi .  \label{eq:Hsol}
\eeq
To obtain the gauge vector in Eq.~\eqref{eq:xi-ansatz} we must also solve $\delta \mathrm{d} \chi = \Box \chi = ( \nabla_{\mu} \zeta^2 ) \nabla_{\nu} H^{\mu \nu}$, that is,
\beq
\Box \chi = \frac{1}{\zetac} \left( \cL_1^\dagger - i a \sin \theta \cD \right) \alpha .
\eeq
This also has an elementary solution,
\beq
\chi = \frac{1}{48 \omega^2} \cD \cD \cL_1^\dagger \cL_2^\dagger \psi .  \label{eq:chisol}
\eeq
In summary, the gauge vector that transforms the radiation-gauge solution \eqref{eq:hrad} to Lorenz gauge via \eqref{eq:hLorenz} is
\beq
\xi^\mu = \zeta^2 \nabla_{\nu} H^{\mu \nu} - g^{\mu \nu} \nabla_\nu \chi ,  \label{eq:gauge-vector-sol}
\eeq
where the key ingredients are in Eqs.~\eqref{eq:Hansatz}, \eqref{eq:Hsol} and \eqref{eq:chisol}.

\textit{Reformulation in terms of GHP calculus.---}
We now rewrite the previous results using the Geroch-Held-Penrose (GHP) formalism \cite{Geroch:1973am} (see Sec.~4.1.1 of Ref.~\cite{Pound:2021qin} for a review). This allows us to: reformulate the results in a compact and coordinate-independent way; eliminate the need for a mode ansatz; and extend the results to the full Kerr-NUT class of Petrov type-D spacetimes. It also allows us to obtain a similar result for the gauge transformation from {\it outgoing} radiation gauge by applying the GHP prime operator along with the identifications $\psi' = \psi^{\rm ORG}$, $\chi' = \chi^{\rm ORG}$ and $H'_{\mu\nu} = H^{\rm ORG}_{\mu\nu}$.
Translating the key ingredients in the gauge transformation to GHP expressions and introducing the Lie derivative, $\LieT$, along the time-translation Killing vector, $T^\mu$, we get
\begin{subequations}
\begin{align}
\label{eq:chi-GHP}
  \LieT^{2} \chi &= -\frac{1}{24}  \edth^2 \bar{\zeta^2} \th^2 \psi, \\
\label{eq:H-GHP} 
  \LieT H_{\mu \nu} &= l_{[\mu} m_{\nu]} \frac{1}{3 \zeta^2} \edth \zetac \th \psi.
\end{align}
\end{subequations}

\textit{Metric perturbation from Weyl scalars.---} We now seek to express the Lorenz-gauge metric perturbation $h_{\mu\nu}^L$ in terms of the Weyl tensor that it generates. In particular, we consider projections $\Psi_0 = C_{lmlm}$ and $\Psi_4 = C_{n\mb n \mb}$ which are invariant under gauge and infinitesimal tetrad transformations. For the metric perturbation \eqref{eq:hrad} or its conjugate, prime, or prime conjugate, one finds after imposing the Teukolsky equation that, respectively, (see e.g.~\cite{Pound:2021qin})
\begin{subequations}
\begin{align}
  \Psi_0 &=
    \frac14 \{0,\,  \th^4 \bar{\psi},\,
     3 M \LieT \zeta^{-4} \psi',\, \edth^4 \bar{\psi}'\}, \\
  \Psi_4 &=
    \frac14 \{-3 M \LieT \zeta^{-4} \psi, \, \edth'^4 \bar{\psi},\, 0,\, \th'^4 \bar{\psi}'\}.
\end{align}
\end{subequations}
If we work with a metric perturbation $h_{\mu \nu} + \bar{h}_{\mu \nu}$ or $h'_{\mu \nu} + \bar{h}'_{\mu \nu}$ alone, then we recover the standard radiation gauge relations between the Hertz potentials and the Weyl scalars \cite{Pound:2021qin}. Alternatively, we can choose the ``antisymmetric'' combination
$h^-_{\mu \nu} = \frac12[h'_{\mu \nu} + \bar{h}'_{\mu \nu} - (h_{\mu \nu} + \bar{h}_{\mu \nu})]$. After imposing the Teukolsky-Starobinsky identities, this leads to the remarkably simple relations \cite{Aksteiner:2016pjt,Aksteiner:2016mol}
\begin{subequations}
\begin{align}
  \Psi_0 = \frac{3M}{4} \LieT \zeta^{-4} \psi', \quad
  \Psi_4 = \frac{3M}{4} \LieT \zeta^{-4} \psi.
\end{align}
\end{subequations}
Note in particular that $\psi$ and $\psi'$ are {\it not} the same as the radiation gauge potentials, and similarly the $h_{\mu\nu}$ and $h'_{\mu\nu}$ appearing in $h^-_{\mu\nu}$ are also different to the radiation gauge metric perturbations.
We can thus reinterpret this as
\begin{equation}
  M \LieT h^-_{\mu \nu} = - \frac{1}{3} \nabla_\beta [\zeta^4 \nabla_\alpha \mathcal{C}_{(\mu}{}^\alpha{}_{\nu)}{}^\beta] + \text{c.c.}
\end{equation}
where $\mathcal{C}^{\mu\alpha\nu\beta} = 4 (\Psi_0 \,  n^{[\mu} \mb^{\alpha]} n^{[\nu} \mb^{\beta]} - \Psi_4 \,  l^{[\mu} m^{\alpha]} l^{[\nu} m^{\beta]})$
is the spin-2 part of the self-dual Weyl tensor with the sign of $\Psi_4$ flipped.
Since $\Psi_0$ and $\Psi_4$ are gauge invariant, these relations also hold after transforming to Lorenz gauge using \eqref{eq:gauge-vector-sol} (or its prime, conjugate, or prime conjugate).

In all three cases, imposing the Teukolsky-Starobinsky identities and the Teukolsky equation reduces four components of the Lorenz-gauge metric perturbation to second-order operators acting on $\Psi_0$ and $\Psi_4$,
\begin{subequations}
\label{eq:h-components-Weyl}
\begin{align}
\label{eq:hll-LG}
  \LieT^{2} h^L_{ll} &= -\frac{1}{3} \big[\zetac^{-2} \edth^2( \zetac^4 \bar{\Psi}_0) + \zeta^{-2} \edth'^2( \zeta^4 \Psi_0)\big], \\
\label{eq:hnn-LG}
  \LieT^{2} h^L_{nn} &= -\frac{1}{3} \big[\zetac^{-2} \edth'^2( \zetac^4 \bar{\Psi}_4) + \zeta^{-2} \edth^2( \zeta^4 \Psi_4)\big], \\
\label{eq:hmm-LG}
  \LieT^{2} h^L_{mm} &= -\frac{1}{3} [\zetac^{-2} \th^2( \zetac^4 \bar{\Psi}_4) + \zeta^{-2} \th'^2( \zeta^4 \Psi_0)], \\
\label{eq:hmbmb-LG}
  \LieT^{2} h^L_{\mb\mb} &= -\frac{1}{3} [\zetac^{-2} \th'^2( \zetac^4 \bar{\Psi}_0) + \zeta^{-2} \th^2( \zeta^4 \Psi_4)].
\end{align}
A fifth component is obtained from the fact that this metric perturbation is traceless,
\begin{equation}
\label{eq:h-LG}
 h = 2(h^L_{m\mb}-h^L_{ln}) = 0.
\end{equation}
\end{subequations}
No such simplification appears possible for the remaining five components, but they can be written in terms of a sixth order operator acting on $\Psi_0$ and $\Psi_4$.

\textit{Spin-$1$ solutions.---}
A set of spin-$1$ solutions satisfying $\Box \xi^\mu = 0$ and $\nabla_\mu \xi^\mu = 0$ were obtained in Ref.~\cite{Dolan:2019hcw,Wardell:2020naz} (see also Ref.~\cite{Frolov:2018ezx,Lunin:2017drx}). They take the form
\begin{align}
\xi^\mu_{(s=1)} = \nabla_\nu \left( \zeta \cH_{(s=1)}^{\nu \mu} \right) + \text{c.c.}, \label{eq:spin1-xi}
\end{align}
where
\begin{align}
\cH^{\mu \nu}_{(s=1)} = 2 \LieT^{-1} \left[ \phi_0 \bar{m}^{[\mu} n^{\nu]} - \phi_2 l^{[\mu} m^{\nu]} \right] .
\end{align}
Here, $\phi_0$ and $\phi_2$ are Maxwell scalars that satisfy the Teukolsky equations for $s=+1$ and $s=-1$, respectively (i.e. $\mathcal{O}\phi_0 = 0 = \mathcal{O}' \phi_2$), and which are linked by the spin-1 Teukolsky-Starobinsky identities. A traceless spin-1 Lorenz-gauge metric perturbation can be constructed from $\xi^\mu_{(s=1)}$ in the now-familiar way, $h^{(s=1)}_{\mu \nu} = -2 \xi^{(s=1)}_{(\mu ; \nu)}$.

\textit{Spin-$0$ solutions.---}
So far, we have only considered trace-free solutions, $h=0$. The trace of the metric perturbation must satisfy
\begin{equation}
  \Box h = 0
\end{equation}
in the homogeneous case by virtue of the contraction of Eq.~\eqref{Lorenz-field-equation}. It is natural to ask: what (non-unique) homogeneous Lorenz-gauge metric perturbation generates a trace $h$?
A suitable metric perturbation is pure-gauge, i.e.,
\beq
  h^{(s=0)}_{\alpha \beta} = -2\xi^{(s=0)}_{(\alpha;\beta)} ,
\eeq
and is generated by a gauge vector that satisfies
\beq
  \nabla_\alpha \xi^\alpha_{(s=0)} = - \frac12 h, \quad \Box \xi^\alpha_{(s=0)} = 0 . \label{eq:xi0-properties}
\eeq
A vector with precisely these properties is
\begin{equation}
  \xi^\alpha_{(s=0)} = \frac12 \LieT^{-1} f^{\alpha \beta} h_{;\beta} + 2 \kappa^{;\alpha}, \label{eq:xi-scalar}
\end{equation}
where
\begin{equation}
    f_{\alpha \beta} = (\zeta + \zetac) n_{[\alpha} l_{\beta]} - (\zeta - \zetac) \mb_{[\alpha} m_{\beta]},
\end{equation}
is the conformal Killing-Yano tensor (we follow here the definition of \cite{Aksteiner:2014zyp}, which differs from that of Ref.~\cite{Frolov:2017kze} by an overall sign), and where $\kappa$ is a scalar field satisfying
\begin{equation}
  \Box \kappa = \frac{1}{2} h .
\end{equation}
It is straightforward to show that the requirements \eqref{eq:xi0-properties} are satisfied by using the properties of the conformal Killing-Yano tensor, namely
\begin{equation}
   f_{\alpha(\beta;\gamma)} = g_{\beta\gamma} T_\alpha - g_{\alpha (\beta} T_{\gamma)}, \;
   f_{\alpha\beta} = f_{[\alpha\beta]},
   \; T^\alpha = \tfrac13 f^{\alpha \beta}{}_{;\beta}.
\end{equation}

In the Schwarzschild case, the two spin-$0$ degrees of freedom, $h$ and $\kappa$, map on to those identified by Berndtson \cite{Berndtson:2007gsc} (see also Khavkine \cite{Khavkine:2020ksv}).

\textit{Completion pieces on Kerr spacetime.---}
In addition to spin-$s$ contributions, the metric perturbation may also contain ``completion'' pieces \cite{Merlin:2016boc,vanDeMeent:2017oet,Aksteiner:2018vze} associated in the Kerr case with infinitesimal changes in the mass and angular momentum of the black hole. Completion pieces are constructed from varying the mass $M$ and specific angular momentum $a=J/M$ parameters, viz.,
\beq
h^{(\partial M)}_{\mu \nu} \equiv \left. \frac{\partial \mathrm{g}_{\mu \nu}}{\partial M} \right|_{a} , \quad h^{(\partial a)}_{\mu \nu} \equiv \left. \frac{\partial \mathrm{g}_{\mu \nu}}{\partial a} \right|_{M} , 
\eeq
where $\mathrm{g}_{\mu \nu}$ is the Kerr metric. Moreover, the conformal mode $h_{\mu \nu}^{(2g)} = 2 \, \mathrm{g}_{\mu \nu}$ automatically satisfies the linearised vacuum field equations. These three pieces are linearly related by the equation
\beq
h_{\mu \nu}^{(2g)} = M h^{(\partial M)}_{\mu \nu} + a \, h^{(\partial a)}_{\mu \nu} + 2 N_{(\mu ; \nu)} ,  \label{eq:completion-relation}
\eeq
with the gauge vector $N^\mu \partial_\mu = t \,\partial_t + r \partial_r$.

Unlike the conformal mode, the perturbations $h^{(\partial M)}_{\mu \nu}$ and $h^{(\partial a)}_{\mu \nu}$ (for $a\neq0$) are {\it not} in Lorenz gauge. To shift to Lorenz gauge, we apply a gauge transformation,
\beq
h_{\mu \nu}^{L(\partial M)} = h_{\mu \nu}^{(\partial M)} - 2 Y_{(\mu ; \nu)} .
\eeq
As $h^{(\partial M)}_{\mu \nu}$  is traceless, it follows that $\Box Y_\mu = \nabla^\nu h_{\mu \nu}^{(\partial M)} = 2 \delta_{\mu}^r / \Delta$. Since the right-hand side is a gradient, the gauge vector is also a gradient, $Y_{\mu} = \nabla_{\mu} y$, and using $\Box ( \nabla_\mu y ) = \nabla_\mu ( \Box y )$, the potential $y$ must satisfy
\beq
\Box y = \int \frac{2}{\Delta} dr = \left(\frac{2}{r_+ - r_-} \right) \ln \left(\frac{r - r_+}{r - r_-}\right) + \text{const}.
\eeq
This equation can be solved by separation of variables. The Lorenz-gauge mode $h_{\mu \nu}^{L(\partial a)}$ follows via Eq.~(\ref{eq:completion-relation}).

The mass and angular momentum content of the $h^{(\partial M)}_{\mu \nu}$ and $h^{(\partial a)}_{\mu \nu}$ modes is assessed by evaluating the conserved charges associated with the background Killing vectors (see Sec.~IIE in Ref.~\cite{Dolan:2012jg}, and Ref.~\cite{Abbott:1981ff}); we find $ Q_{(t)} = 1, Q_{(\phi)} = - a$ and $Q_{(t)} = 0, Q_{(\phi)} = - M$, respectively.

\textit{Discussion.---}
We have obtained a set of Lorenz-gauge metric perturbations which satisfy the vacuum field equations [Eq.~(\ref{Lorenz-field-equation}) with $T_{\mu \nu} = 0$]. In the frequency domain, the spin-$0$, spin-$1$ and spin-$2$ metric perturbations can be expressed in terms of separable modes, that is, radial and angular functions ${}_sR_{\ell m \omega}(r)$ and ${}_sS_{\ell m \omega }(\theta)$  satisfying the vacuum Teukolsky equations for $s=0$, $s=\pm 1$ and $s=\pm 2$. It is notable that, although the construction of the spin-2 modes starts with the radiation-gauge potentials $\psi$, the Lorenz-gauge metric components in Eq.~\eqref{eq:h-components-Weyl} can be written in terms of Weyl scalars only, without reference to $\psi$. We also note however that it is likely that the zero frequency modes of the spin-$2$ case will need to be treated separately, as has been done for the spin-$1$ case \cite{Wardell:2020naz}.

Several extensions of this work suggest themselves. First, extending the Lorenz-gauge formalism to include source terms ($T_{\mu \nu} \neq 0$). Second, constructing solutions for GSF particle-inspiral scenarios by demanding global regularity (in vacuum regions) on a metric perturbation constructed from a sum over a complete set of vacuum modes. Third, the application of these Lorenz-gauge solutions in second-order GSF applications \cite{Pound:2019lzj,Warburton:2021kwk,Wardell:2021fyy}, ultimately leading to the production of waveforms for extreme mass ratio systems with a spinning primary (larger) black hole.

\begin{acknowledgments}
\textit{Acknowledgements.---}
With thanks to Leanne Durkan, Vahid Toomani, Stephen Green, Stefan Hollands, Adam Pound, Leor Barack, Adrian Ottewill, Amos Ori, Saul Teukolsky, Bernard Whiting and Lars Andersson for discussions. Many of the calculations in this work were enabled by the \textsc{xAct} \cite{xTensor,xTensorOnline} tensor algebra package for \textsc{Mathematica}.
S.D.~acknowledges financial support from the Science and Technology Facilities Council (STFC) under Grant No.~ST/P000800/1, and from the European Union's Horizon 2020 research and innovation programme under the H2020-MSCA-RISE-2017 Grant No.~FunFiCO-777740.  
\end{acknowledgments}

\bibliographystyle{apsrev4-1}
\bibliography{LorenzGaugeKerr}

\end{document}